# A SELF-CONSISTENT MODEL OF THE CIRCUMSTELLAR DEBRIS CREATED BY A GIANT HYPERVELOCITY IMPACT IN THE HD172555 SYSTEM

B.C. Johnson[1], C.M. Lisse[2], C.H. Chen[3], H. J. Melosh[4,1], M.C. Wyatt[5], P. Thebault[6], W.G. Henning[7], E. Gaidos[8], L.T. Elkins-Tanton[9], J.C. Bridges[10], A. Morlok[11]

[1] Department of Physics, Purdue University, 525 Northwestern Avenue, West Lafayette, IN 47907, USA

johns477@purdue.edu

[2] JHU-APL, 11100 Johns Hopkins Road, Laurel, MD 20723, USA

[3] STScI 3700 San Martin Dr. Baltimore, MD 21218, USA

[4] Department of Earth, Atmospheric, and Planetary Sciences, Purdue University, West Lafayette, IN 47907, USA

[5] Institute of Astronomy, University of Cambridge, Madingley Road, Cambridge, CB3 0HA, UK

[6] LESIA, Observatoire de Paris, F-92195 Meudon Principal Cedex, France

[7] NASA Goddard Space Flight Center, 8800 Greenbelt Rd., Greenbelt, MD 20771

[8] Department of Geology and Geophysics, University of Hawaii at Manoa, Honolulu, Hawaii 96822, USA

[9] Department of Terrestrial Magnetism, Carnegie Institution for Science, Washington DC 20015, USA

[10] Department of Physics and Astronomy, University of Leicester, LE1 7RH, UK

[11] Department of Physical Sciences, The Open University, Walton Hall, MK7 6AA Milton Keynes, UK

**Keywords:** astrochemistry; infrared: stars; radiation mechanisms: thermal; techniques: spectroscopic; stars: planetary systems: formation; protoplanetary disks

53 Pages, 7 Figures, 0 Table

Please address all future correspondence, reviews, proofs, etc. to:

Brandon C. Johnson

Department of Physics

Purdue University

525 Northwestern Avenue

West Lafayette, IN 47907

Johns477@purdue.edu

## Abstract:

Spectral modeling of the large infrared excess in the Spitzer IRS spectra of HD 172555 suggests that there is more than $10^{19}$ kg of sub-micron dust in the system. Using physical arguments and constraints from observations, we rule out the possibility of the infrared excess being created by a magma ocean planet or a circumplanetary disk or torus. We show that the infrared excess is consistent with a circumstellar debris disk or torus, located at ~ 6 AU, that was created by a planetary scale hypervelocity impact. We find that radiation pressure should remove submicron dust from the debris disk in less than one year. However, the system's mid-infrared photometric flux, dominated by submicron grains, has been stable within 4% over the last 27 years, from IRAS (1983) to WISE (2010). Our new spectral modeling work and calculations of the radiation pressure on fine dust in HD 172555 provide a self-consistent explanation for this apparent contradiction. We also explore the unconfirmed claim that ~$10^{47}$ molecules of SiO vapor are needed to explain an emission feature at ~8 μm in the Spitzer IRS spectrum of HD 172555. We find that unless there are ~$10^{48}$ atoms or 0.05 $M_{\oplus}$ of atomic Si and O vapor in the system, SiO vapor should be destroyed by photo-dissociation in less than 0.2 years. We argue that a second plausible explanation for the ~8 μm feature can be emission from solid SiO, which naturally occurs in submicron silicate "smokes" created by quickly condensing vaporized silicate.

## 1 – Introduction

According to the most widely accepted theories of planet formation (Ida & Lin 2004; Chambers 2004; Kenyon & Bromley 2004), the planets we see today in the solar system are the final winners of a long series of violent impacts, mergers, and accretion events. In the very beginning of a system, grains of heavy elements condense out of gaseous circumstellar disks and into dust grains. Collisions and aggregational clumping result in the accretion of these grains into solid bodies a few centimeters in diameter. Collective effects such as gravitational instabilities then lead to the production of planetesimals a kilometer or more in diameter (Blum 2010). Once km-sized planetesimals are common, continued accretion yields protoplanets (often called embryos or planetary cores) on timescales of $10^5$ to $10^6$ years, which continue to collide and merge into full-sized planets. One or more planets in a system may grow to about 10 times the mass of Earth, enough to capture large amounts of gas as they sweep through the nebula, creating deep atmospheres like those of Jupiter and Saturn. The primordial phase of disk evolution, when giant planets form, ends with the dissipation of all disk gas content. For most stars this process seems to require 1 to 6 Myr, with 3 Myr often cited as a typical lifetime for gaseous disks (Pascucci *et al.* 2006, Hernandez *et al.* 2007, Currie *et al.* 2009). Over the next few Myr, the remnant primordial circumstellar dust is either blown out by the flux of stellar radiation, decelerated by Poynting Robertson drag and accreted onto the primary, destroyed in collisions, or accreted onto growing rocky planetesimals. Throughout the next ~30 Myr, rocky terrestrial protoplanets continue to accumulate planetesimals near the primary and icy planets accrete icy planetesimals outside the system's ice line. By 100 Myr, current models have solar systems at a relatively mature stage, with little accretional growth occurring (although significant surface evolution and alteration can occur via impacts and deposition of a late veneer). Over the next Gyrs of the primary star's main sequence life, observable amounts of dust in a debris disk may be created by the steady state grinding of planetesimals or the occasional large impact.

Thus, large bodies in solar systems grow and evolve via collisional accretion and fragmentation. However, as much as we may want to understand the processes that created present day solar system structures, evolutionary processing over the last 4.5 Gyr has erased much of the evidence. Fortunately, we can learn about these processes by observing nearby young stars forming their own planetary systems, because the same physical processes that formed the bodies & debris clouds in the solar system are operating in these systems. By finding exo-analogs for solar system structures such as the collisionally active Kuiper Belts in resolved circumstellar disks (e.g., η Corvi, Wyatt *et al.* 2005, Lisse *et al.* 2012); the dense warm zodiacal clouds around stars such as HD 69830, created by asteroid belt collisions (Beichman *et al.* 2005, Lisse *et al.* 2007); or the massive collisional aggregation occurring in the HD 113766 system (Lisse et al. 2008), these systems can help us learn more about the physics causing these processes to occur, what they create, and how they evolve (Wyatt *et al.* 2007, Kriviov 2010).

Of particular interest for this work, we note that in the lunar formation event, the incoming Mars sized planetesimal Theia must have impacted at a relative speed about equal to the escape velocity for the Earth-Moon system ($v_{escape}$ = 12 km sec$^{-1}$) (Canup 2004, 2008). Jackson and Wyatt (2012) estimate that in addition to creating a substantial circumplanetary disk that subsequently forms the Moon, this impact would create an ~10$^{23}$ kg underlined{circumstellar} debris disk, which could be readily detectable around other stars by Spitzer 24 μm surveys for ~25 Myr after the initial impact. As we show in **Section 2**, the Spitzer IRS spectra, imaging observations, and the interferometric visibility of HD 172555 are consistent with emission from a circumstellar debris disk or torus located at ~6 AU from the primary created by a planetary scale hypervelocity impact. In **Section 2** we also introduce several standing questions about the HD 172555 debris disk. Although submicron dust should be quickly removed from the system by radiation pressure blowout, observations show that the system's mid-infrared photometric flux, which is dominated by emission from submicron dust, has remained constant within 4% over the last 27 years (**Figure 1**). We also examine the problems associated with the as yet unconfirmed suggestion, made by Lisse et al. 2009, that there are $10^{47}$ molecules of SiO vapor in HD 172555. These problems include a poorly constrained condensation lifetime of SiO as well as an unknown photo-dissociation lifetime of SiO.

In **Section 3** we consider the stability of fine dust in HD 172555. First we calculate the lifetime of grains in HD 172555 due to radiation pressure blowout and Poynting Robertson drag (**Section 3.1**). Then we show that it is possible to construct a particle size distribution that is consistent with our radiation pressure blowout calculations and the Spitzer IRS spectra of HD 172555 (**Section 3.2**). In **Section 4** we provide constraints on the possible presence of SiO vapor in HD 172555. We show that it could take $10^3$-$10^4$ yr for SiO vapor to condense onto existing grains (**Section 4.2**). We then argue that, unless there are ~$10^{48}$ atoms or 0.05 $M_\oplus$ of Si and O vapor in the system, SiO should be destroyed by photo-dissociation in less than 0.2 yr (**Section 4.3 and Section 4.4**). Finally, in **Section 5** we introduce the alternative possibility that the ~8 μm feature observed by Spitzer is emission from solid SiO rather than SiO in the gas phase.

## 2 – HD 172555: Discovery of a Silica-Rich Debris Disk

Dust grains orbiting a star absorb visible light from the primary and re-emit mainly in the infrared, causing systems with abundant dust to have excess emission in the infrared (Backman 1993). In 2006, Chen et al. (hereafter Ch2006) published a Spitzer IRS 5-35 μm survey of 59 of the brightest IRAS sources with large infrared excesses attributed to circumstellar dust. Of their 59 systems, 5 showed unusually strong, high contrast warm dust features: HD 113766A, η Corvi, η Tel, HD 172555, and HR 3927. Ch2006 showed that they were able to explain the observed excess emission spectra using a mix of crystalline and amorphous silicates and silica, and larger "rubble" emitting as blackbodies. In 2009, as part of a detailed follow-up study of the most interesting Ch2006 sources, Lisse *et al.* (2009) (hereafter Li2009) utilized absorption and emission spectra of laboratory dust analogues coupled with lessons learned from modeling Deep Impact and STARDUST cometary dust to identify HD 172555 as exhibiting a highly unusual composition compared to the other IRAS-bright systems. The infrared excess of HD 172555, a 12 Myr old A5V star in the β Pictoris moving group, is dominated by emission from dust with temperature ~340 K and a strong emission peak at ~9 μm (**Figure 1, Figure 2**), This peak at 9 μm is attributed to the Si-O bond vibration of amorphous silica and, along with a secondary 20 μm feature, is quite distinctive and unlike any known olivine or pyroxene emission spectra (**Figure 2**, Li2009).

Modeling the Spitzer IRS spectrum of HD 172555 using a mix of submicron amorphous silica, silicate, and metal sulfide dust particles, SiO gas, and large optically thick dust particles, Li2009 produced a greater than 95% confidence level match to the spectrum. They found some $6 \times 10^{19} - 2 \times 10^{20}$ kg of "fine" 0.1-1000 μm dust, and at least $3 \times 10^{21} - 1 \times 10^{22}$ kg of large, optically thick "rubble" emitting as blackbodies. Using both the maximum temperature of the fine dust (as calibrated versus the Deep Impact experiment, Lisse *et al.* 2006) and the effective temperature of the large rubble, ~200 K, they placed the dust populations at 5.8 ± 0.6 (1σ) AU from the system's A5V primary star. Ruling out other possible formation mechanisms, such as shocks, stellar flares, or β Meteoroids, Li2009 concluded that the glassy silica must have been formed in a giant hypervelocity impact, with an impact velocity greater than 10 km s$^{-1}$, between 2 large (greater than $M_{Mercury}$) rocky, silicate-rich bodies at roughly 6 AU from the central star. In addition to an abundance of glassy silica, the Spitzer spectrum also has a strong residual feature centered at ~8 µm. Li2009 found the fundamental rotational-vibrational linear stretch complex of the SiO vapor molecule was a good match to the residuals. (This feature cannot be seen in **Figure 1**, because photospheric emission from HD 172555 has not been removed, but as shown in Li2009, when the photospheric emission is removed, an 0.07 Jy excess is observed at 8 µm.) Without the inclusion of SiO vapor, the Spectral model only corresponds to 0.03 Jy at 8 µm, an error of ~70% (Li2009). Li2009 found that if the excess emission at 8 µm is due to SiO vapor fluorescing at ~6 AU from the primary, the emission is produced by ~$10^{22}$ kg or $10^{47}$ molecules of optically thin SiO vapor. This finding further supported the conclusion that the HD 172555 debris disk was created by a massive hypervelocity impact, as SiO vapor is expected to be produced via the dissociation of silicates (e.g., olivines and pyroxenes) following vaporization.

The Spitzer observations of HD 172555 are spatially unresolved, and analysis to date has relied only on spectral decomposition. Since 2009, however, there have been several attempts to image the HD 172555 debris disk that have given us spatial constraints on the location of the dust. An observation, recently reported by Pantin & Di Folco (2011), using a so-called "lucky imaging" VLT/VISIR 8-13 µm photometric imaging technique, is consistent with an axisymmetric circumstellar torus located at ~6 AU. The 18 µm emission was also resolved using TReCS in Smith et al. (2012) suggesting emission located at ~8 AU, consistent with the ~6 AU results of Pantin & Di Folco (2011) and Li2009. Conversely, the 12 µm emission was not resolved by TReCS. Inability to resolve the 12 µm emission, combined with N-band interferometric visibilities from MIDI on VLTI, led Smith et al. (2012) to conclude that warm dust emitting at 10-12 µm is an axisymmetric circumstellar disk located somewhere within the 1-8 AU region. ***All three observations argue against the possibility of an asymmetric or point source structure for the observed excess infrared emission.***

From the presence of meta-stable amorphous silica and SiO gas, Li2009 argued that the HD 172555 circumstellar debris cannot be in thermodynamic equilibrium. Further work focusing on the dynamics of the HD 172555 debris disk has raised questions regarding the temporal stability of the system (Lisse *et al.* 2010). E.g., it is not clear that fine dust on the order of ~1 µm can be dynamically stable, against radiative blowout, in the radiation field of the HD 172555. Yet there must be copious fine dust to produce the strong emission features seen in the mid-infrared. If submicron dust is not dynamically stable something must be replacing it in a roughly constant fashion. Additionally, the radiation field of this A5V star should produce copious amounts of UV photons, shortward of 1500 Å, capable of photolytically destroying the SiO molecule, which has a molecular binding energy of 8.26 eV (Tarafdar and Dalgarno 1990). Either SiO vapor is being actively created or some other species is creating the observed 8 µm feature. Determining the SiO lifetime against photo-dissociation or re-condensation into the solid state is thus of great importance. Finally, the best fit particle size distribution (PSD) found by Li2009 using their spectral decomposition technique has a very steep number density per unit grain size, $dn/da \propto a^{-3.95 \pm 0.10}$, where $a$ is the particle radius. This slope is much steeper than canonical steady state collisional distributions, which is given approximately by $dn/da \propto a^{-3.5}$ for particles with dispersal thresholds that are independent of particles size (Backman and Paresce 1993). Although the system's PSD implies that there is abundant submicron dust in the system, which should be removed from the system by radiation pressure, the system's mid-infrared photometric flux has stayed fairly stable, within 4%, over 27 years, from IRAS (1983) to WISE (2010) (**Figure 1**).

Given our current understanding of solar system formation and evolution, we investigate a number of possible physical scenarios, including circumplanetary disks and magma-ocean covered planets, to answer these stability concerns by creating dynamically and photolytically stable collections of silica and SiO gas in HD 172555. We only consider major impact driven events or high temperature outflows, as the amount of dust mass involved to emit the Jy-level infrared emission detected by Spitzer at a distance of 29 pc is at least $10^{21}$ kg, a mass consistent with a Ceres sized body (Li2009). However, on physical and observational grounds, we have ruled out physical scenarios involving circumplanetary disks and magma-ocean covered planets. Thus, we conclude the infrared excess of HD 172555 is produced by a circumstellar disk or torus created by an initial giant impact between two planet-sized bodies. In the following sections we explain the stability of submicron dust in HD 172555 and determine the conditions necessary to keep SiO vapor present in the system. For the sake of completeness, the models which could have plausibly produced large amounts of stable silica and SiO gas, but that we have ruled out given the current set of observational data for HD 172555, are described in Appendix A.

## 3 – Stability of Fine Dust

In **Section 2** we introduced several questions about the Li2009 study. Li2009 found that a steeper than collisional PSD with a $dn/da \propto a^{-3.95\pm0.10}$ is required to fit the Spitzer IRS spectrum of HD 172555. This PSD corresponds to a system with mid-infrared emission dominated by submicron grains, which should be quickly removed from the system by radiation pressure. This quick removal of submicron grains seems to be at odds with the apparent stability of the system's observed mid-infrared photometric flux over 27 years (**Figure 1**). However, as we show in the following sections, it is possible to construct a PSD that is both long-lived and consistent with the Spitzer IRS spectrum. In **Section 3.1** we calculate the lifetime of grains in HD 172555 due to radiation pressure and Poynting Robertson drag. Our calculations show that glassy silica grains between 0.02 µm and 1.5 µm in radius should be removed from the debris torus in less than one year. In **Section 3.2,** based on the laboratory work of Takasawa *et al.* (2011), we argue that a steeper than collisional PSD is consistent with dust being sourced by recent hypervelocity grinding. We also show that it is possible to construct a PSD that is consistent with the calculation of dust that will be blown out by radiation pressure and the Spitzer IRS spectra of HD 172555 (**Figure 6 bottom**).

## 3.1 – Radiation Pressure Blowout

To estimate the stability of fine dust we calculate the ratio of the radiation pressure force to gravitational force as a function of particle size

$$\beta(a) = \frac{3L_\star \langle Q_{pr}(a) \rangle}{16\pi G M_\star ca\rho}, \qquad (1)$$

where $L_\star$ and $M_\star$ are the luminosity and mass of HD 172555, $c$ is the velocity of light, $\rho$ is the density of the grains, $a$ is the radius of the grains, and $\langle Q_{pr}(a) \rangle$ is the radiation pressure coefficient for particles of radius $a$ averaged over the stellar spectrum (Artymowicz 1988).

β Pictoris, a fellow 12 Myr old A5V member of the β Pictoris moving group (BPMG), has the same effective temperature $T_{eff}$ = 8000 K as HD 172555. Thus, for particles of the same composition and density, the $\beta$ values for HD 172555 are given by the following scaling $\beta(a) = \beta(a)_{\beta_{pic}}(M_{\beta_{pic}}/M_\star)\,(L_\star/L_{\beta_{pic}})$ , where $L_{\beta_{pic}} = 6.5\,L_\odot$ , $M_{\beta_{pic}} = 1.7 M_\odot$ , and $\beta(a)_{\beta_{pic}}$ are the luminosity, mass, and β values for βPictoris (Artymowicz 1988), and $L_\star = 9.5\,L_\odot$ and $M_\star = 2 M_\odot$ are the luminosity and mass of HD 172555 (Li2009). The calculated values for $\beta(a)$ in HD 172555 are shown in **Figure 3**.

We can use the value of Beta for a dust particle to determine its long-term dynamical stability versus the forces of stellar radiation pressure and gravity. For example, dust with $\beta < 1/2$, produced from parent bodies on circular orbits having $\beta = 0$, will have stable orbits. On the other hand, when $\beta > 1/2$, the dust, produced from parent bodies on circular orbits having $\beta = 0$, will move outward with a terminal radial velocity

$$v_r \approx \left[\left(\frac{2GM_\star}{R}\right)\left(\beta - \frac{1}{2}\right)\right]^{\frac{1}{2}} \qquad\qquad (2)$$

(Su *et al.* 2005). For HD 172555, $R \sim 6$ AU is the radius at which the dust is created, $G$ is the gravitational constant, and $M_\star = 2 M_\odot$ (Li2009).

Using the optical properties of obsidian, a majority amorphous silica species with ~75% $SiO_2$ and ~14% $Al_2O_3$ by mole (Artymowicz 1988 and references therein), a composition that is comparable to the ~69% $SiO_2$, 10% MgO, and 9% $Al_2O_3$ mixture of Bediasite and obsidian found for the "fine dust" in HD172555 (Li2009), **Figure 3** shows that dust between 0.02 µm and 1.5 µm in radius will be removed from HD 172555 by radiation pressure. To estimate the rate at which dust is removed from the torus we consider the crossing time,

$$\tau_c \approx \frac{r}{v_r}, \qquad\qquad (3)$$

where $r$ is the minor radius of the torus. A best-fit model of the 18 µm resolved emission gives a disk with a width $dr \approx 9.5$ AU (For detailed description the debris geometry see **Figure 4**). Assuming that the disk can be approximated by a torus, the minor radius of the torus is about half of the disk's width or $r \approx 4.75$ AU (Smith *et al.* 2012). Smith et al. (2012) estimate that the lower limit for disk width is $dr \approx 1$ AU or $r \approx 0.5$ AU. Additionally, the surface area of the dust requires $r > 0.01 /\sin(i)$ AU, where $i$ is the inclination of the disk or torus (Li2009). If $r = 0.01 /\sin(i)$ AU, the torus would be optically thick and emit as a blackbody. The Spitzer IRS spectrum is not consistent with blackbody emission, therefore, the torus must be optically thin with $r$ significantly greater than $r = 0.01 /\sin(i)$ AU. Thus, we assume that $r$ is on the order of 1 AU and leave it as a variable in all of our calculations. The imaging observations also show that mid-infrared emission comes dominantly from a limited spatial region described by a disk or torus and emission from dust outside of this torus or disk produces a negligible portion of the observed infrared excess (Pantin & Di Folco 2011, Smith *et al.* 2012). Thus, when assessing the photometric stability of the system we can safely neglect any emission from dust that is blown out of the torus.

We find that dust between 0.02 µm – 1.5 µm in radius will be removed from the torus in $\tau_c < 0.4\,(r/AU)$ yr (**Figure 5**). The lifetime of submicron dust, $\tau_c < 0.4\,(r/AU)$ yr seems inconsistent with observational evidence that the infrared excess of HD 172555, which is dominated by emission from submicron dust, has appeared photometrically stable over the ~27 years from its IRAS discovery in 1983 to its re-observation by WISE in 2010 (**Figure 1**), let alone its stability on the Myr timescales of the system's age and the estimated circumstellar disk lifetime calculated by Jackson & Wyatt (2012). However, we show next that it is possible to construct a PSD that is consistent with the calculation of dust that will be blown out by radiation pressure and the Spitzer IRS spectra of HD 172555 (**Figure 6 bottom**).

## 3.2 – Particle Size Distributions

New spectral modeling work, using the same dust composition found in Li2009 and the same methods of Li2009, shows that additional PSDs are capable of giving a better than 95% confidence level fit to the Spitzer IRS spectrum of HD 172555 (**Figure 6 top**). The mid-infrared spectral modeling is degenerate in that any collection of dust particles that satisfies the following integral equation will produce emission that is essentially identical to the observed Spitzer spectrum. We have

$$F_\lambda = \frac{1}{\Delta^2} \sum_i \int_0^\infty B_\lambda\big(T_i(a,R)\big)\, Q_{abs,i}(a,\lambda)\, \pi a^2 \frac{dn_i(R)}{da}\, da \qquad (4)$$

where the subscript denotes each individual dust component of the ith composition (e.g. olivine, silica,etc.), $T_i(a,R)$ is the particle temperature for a particle of radius $a$ and composition $i$ at a distance R from the primary, $\Delta$ is the distance from Spitzer to the dust, $B_\lambda$ is the blackbody radiance at wavelength $\lambda$, $Q_{abs,i}(a,\lambda)$ is the emission efficiency of the particle of composition $i$ at wavelength $\lambda$ and particle size $a$, $dn/da$ is the differential PSD, the integral is over all sizes, and the sum is over all compositions. In practice, for a spectrum with strong infrared emission features, the compositional mix of materials is relatively uniquely defined. The overall run of temperature with particle size is as well, which helps specify the location of the dust with respect to the central star. The size distribution of the dust can have some play or flexibility in it; a population of 1-10 um silicaceous dust, for example, will produce both strong features and some mid-infrared continuum; a population of extremely small (< 1 μm) particles, which produces only strong emission features plus a population of extremely large (>100 μm) particles, which produces only continuum, may produce similar model infrared spectra, and depending on the quality of the observed spectrum, it may be difficult to distinguish between the two.

The non-unique nature of the spectral modeling allows us to test additional PSDs that can potentially yield a better than 95% confidence level fit to the Spitzer IRS spectrum (as did original PSD published by Li2009; **Figure 6**). For a single power law PSD, the best-fit PSD is $dn/da \propto a^{-3.95 \pm 0.10}$ (this is the same PSD published by Li2009; **Figure 6 top**). **Figure 6 middle** shows a broken power law, with $dn/da \propto a^{-3.55}$ below 1 μm and $dn/da \propto a^{-4.02}$ above 1 μm. Motivated by calculation of $\beta$ for obsidian grains, we show that a PSD with $dn/da \propto a^{-3.95}$ with a reduction in the number of grains between 0.1 μm and 1 μm is also viable (**Figure 6 bottom**).

Although the three PSD's in **Figure 6** look very different, they have many similarities. All of the PSDs are very steep with approximately $dn/da \propto a^{-3.95 \pm 0.10}$ compared to the canonical steady state distribution given approximately by $dn/da \propto a^{-3.5}$ for particles with dispersal thresholds that are independent of particles size (Backman and Paresce 1993). Although the PSD's have particles up to 100 μm in them, these steep PSDs have most of their mass contained in the small particles and all of these PSDs correspond to $10^{19}$-$10^{20}$ kg of submicron dust. After removal of the emission created by the PSDs there is still a large 200 K blackbody signature as described in Li2009 (**Figure 2**). Following Li2009 and assuming that this emission is created by the smallest particles capable of emitting as blackbodies ~100 μm, all three PSDs require the addition of $10^{21} - 10^{22}$ kg of large "rubble". Following Li2009 we call the warm 340 K dust described by the PSDs "fine dust" and the larger particles emitting as 200K blackbodies, which are not included in the PSDs, "rubble". The temperature difference between the "rubble" and "fine dust" is expected, because the large optically thick "rubble" particles co-located with the "fine dust" are much better at radiating heat and needs only to reach a color temperature of ~ 200 K to establish energy balance with the incident starlight.

Lab work by Takasawa *et al.* (2011) shows that a much steeper-than-collisional size distribution is consistent with PSDs created by hypervelocity impacts. Takasawa *et al.* (2011) measured the size distribution for seven hypervelocity shots with impact velocities ranging from 9-61 km/s. The average slope of the PSDs for the seven shots gives $dn/da \propto a^{-4.4\pm0.8}$ (Takasawa *et al.* 2011). Although they use a range of impact velocities, there is no obvious correlation between the slope of the PSD and the impact velocity.

We believe the steeper than collisional PSDs found for hypervelocity impacts in the laboratory indicates that the steeper than collisional PSDs derived for the "fine dust" in HD 172555 by Li2009 implies that most of the "fine dust" around HD 172555 is sourced by hypervelocity impacts. However, we do not believe this "fine dust" was created by the initial planetary scale hypervelocity impact, but by subsequent hypervelocity impacts between sub-millimeter-sized particles similar to those used by Takasawa et al. (2011). Other support for a constant hypervelocity grinding scenario is produced by dynamical modeling of the <u>circumstellar</u> debris disk created by the Moon forming impact, which shows that typical collision velocities in the <u>circumstellar</u> disk are $\geq$ 5 km s$^{-1}$ (Jackson and Wyatt 2012). Although this velocity depends on the details of the orbital elements of the debris, the study of Jackson and Wyatt (2012) shows that hypervelocity grinding is not unprecedented.

In addition to being sourced by hypervelocity impacts, these steep PSDs also require that the particles have not had time to become a steady state collisional distribution. The time it takes for particles to relax to the steady state collisional distribution is their collisional lifetime. Following Wyatt et al. (1999) we find the collisional lifetime of the smallest particles that dominate the surface area of the torus is given by

$$\tau_{col} = \frac{t_{orb}}{4\pi f_A}, \qquad (5)$$

where $t_{orb} \sim 10$ yr is the orbital timescale at 6 AU, and $f_A = A_{dust}/2\pi R 2r$ is the fractional cross sectional area of the torus that the particles occupy. The cross sectional area of the dust is $A_{dust} = (1-5) \times 10^{21}$ m$^2$ (Li2009). Combining this we find

$$\tau_{col} = (300 - 1400)\left(\frac{r}{AU}\right) \text{ yr}. \qquad (6)$$

To observe a steep PSD with $dn/da \propto a^{-3.95 \pm 0.10}$ the fine dust must have only been created within a collisional timescale. As previously stated, the canonical steady state distribution with $dn/da \propto a^{-3.5}$ is for particles with dispersal thresholds that are independent of particle size (Backman and Paresce 1993). The dispersal threshold, $Q_D$, is the specific incident energy required to cause a catastrophic collision, where a catastrophic collision is one where the mass of the largest object remaining after a collision is half the mass of the largest object before the impact. It is possible that we are observing a steady state PSD if the dispersal threshold $Q_D \propto a^{-0.92}$ (Wyatt et al. 2011). This dependence is significantly steeper than typical dispersal threshold size dependences with approximately $Q_D \propto a^{-0.4}$ (Benz and Asphaug 1999).

Based on the $\beta$ values from **Figure 3,** we expect that the PSD agreement with laboratory studies should only apply to dust larger than ~1 μm, as the submicron dust is still removed on a time scale of $\tau_c < 0.4 \ (r/AU)$ yr by radiation pressure. Thus the $dn/da \propto a^{-3.95}$ PSD (**Figure 6 top**) corresponds to a system in which the submicron dust is somehow bound to the system and long lived. If the debris torus is optically thick enough to reduce the effective luminosity of HD 172555 from $L_\star = 9.5 \ L_\odot$ to $L_{eff} \approx L_\odot$ within the torus, then β values for obsidian (**Figure 3**) would be ~10 times smaller. This would make $\beta < 0.5$ for dust of all sizes meaning no radiation pressure blowout would occur. Without any reason to believe the debris torus would be optically thick at the peak wavelength in the HD 172555 spectral energy distribution, we conclude that this situation is unlikely.

The PSD with $dn/da \propto a^{-3.95}$ and a reduction in the number of grains between 0.1 μm and 1 μm (**Figure 6 bottom**), reconciles the apparent contradiction of mid-infrared emission that is photometrically stable over 27 years (**Figure 1**) and dust between $0.02\,\mu\text{m} - 1.5\,\mu\text{m}$ in radius being removed from the torus in $\tau_c < 0.4\,(r/AU)$ yr by radiation pressure (**Figure 5**). As previously discussed this PSD provides a better than 95% confidence level fit with the Spitzer IRS spectrum and should remain stable over a lifetime equal to the collisional lifetime of the torus, which is determined by the lifetime of the largest object in the debris disk and could be longer than 25 Myr (Jackson and Wyatt 2012). For obsidian, **Figure 3** illustrates that $\beta > 0.5$ for particles between 0.02 μm and 1.5 μm and thus they should be quickly ejected from the system by radiation pressure. We argue that the difference between calculated size of the blowout grains (**Figure 3**) of 0.02 μm - 1.5 μm and the blowout grain size of 0.1 μm -1 μm obtained from the spectral modeling (**Figure 6 bottom**) are trivial considering the uncertainties associated with each calculation.

The calculation of $\beta$ assumes the particles are ~75% $SiO_2$ and ~14% $Al_2O_3$ by mole (Artymowicz 1988 and references therein) while the spectral modeling of Li2009 shows that the "fine dust" is dominated by a combination of Bediasite and obsidian with an average composition that is ~69% $SiO_2$, 10% $MgO$, and 9% $Al_2O_3$. Although we have neglected species with mole percent less than 5%, this comparison shows that the obsidian used in **Figure 3** is only an approximate match to the "fine dust" in HD 172555. As shown in equation 1, $\beta \propto 1/\rho$, where $\rho$ is the bulk density of the grains, so if $\rho$ is larger than the 2370 kg m$^{-3}$ used by Artymowicz (1988), $\beta$ could be somewhat smaller than the values shown in **Figure 3**. Another source of uncertainty comes from the degeneracy associated with the PSDs and spectral modeling described at the beginning of this section. Although, the submicron obsidian grains with $\beta < 0.5$ will spiral into the primary in $10^4$-$10^6$ yr due to Poynting Robertson drag, as long as the fine dust is replenished by collisions, the emission will remain stable over a lifetime equal to the collisional lifetime of the torus, which could be longer than 25 Myr (Jackson and Wyatt 2012).

## 4 –Constraints on Possible Presence of SiO Vapor

The claim, made by Li2009, that there are $10^{47}$ molecules of SiO vapor in HD 172555 is currently unconfirmed. In this section we explore the conditions necessary to keep $10^{47}$ molecules of SiO vapor in HD 172555 over reasonable timescales. In **Section 4.1** we show that if SiO vapor is present in the system, the vapor is not in equilibrium with the dust. Then in **Section 4.2** we show that it could take $10^3$-$10^4$ yr for SiO vapor to condense onto existing grains. In **Section 4.3** we estimate the photo-dissociation rate of SiO and in **Section 4.4** we consider the chemistry of a self-shielded torus where SiO is created at the same rate it is destroyed by photolysis. Showing that, unless there is ~$10^{48}$ atoms or 0.05 $M_\oplus$ of Si and O vapor in the system, SiO should be destroyed by photo-dissociation in less than 0.2 yr. Finally, in **Section 5** we introduce the possibility that Spitzer has observed emission from solid SiO rather than SiO in the gas phase.

## 4.1 – Equilibrium of SiO Vapor-Solid System

Assuming the SiO vapor is in thermal equilibrium with the $SiO_2$ dust, we can estimate the equilibrium vapor pressure of SiO above solid $SiO_x$ at a temperature of ~340 K, the approximate temperature of the fine silica dust found by Li2009. We make the estimate using an empirical formula derived by Ferguson and Nuth (2008) for the equilibrium vapor pressure of SiO above solid SiO$_x$ with a stoichiometric x=1,

$$\log_{10}(P_{SiO}/Pa) = 13.29 \pm 0.39 - (17740 \pm 550)/(T/K) \qquad (7)$$

At 340 K the formula gives $P_{SiO} \sim 10^{-39}$Pa. If we use $T = 200$ K, the equilibrium vapor pressure is even lower, $P_{SiO} \sim 10^{-75}$Pa. To convert pressure to number of molecules, we assume that SiO vapor is the only vapor present in the system and treat it as an ideal gas with pressure,

$$P_{SiO} \approx \frac{N_{SiO}k_bT}{V_{torus}}, \qquad (8)$$

where $N_{SiO}$ the number of SiO vapor molecules $k_b$ is Boltzmann's constant. Assuming the SiO vapor is in equilibrium with the fine dust, it will have a temperature $T = 340$ K. Lastly, we assume the SiO vapor is co-spatial with the dust and in a torus with a volume

$$V_{torus} = 2\pi R \times \pi r^2 \approx 4 \times 10^{35} \left(\frac{r}{AU}\right)^2 m^3, \qquad (9)$$

where $R \approx 6AU$ is the major radius of the torus and $r$, the minor radius of the torus. This leads to

$$N_{SiO} \approx 10^{17} \left(\frac{r}{AU}\right)^2 molecules, \qquad (10)$$

which is 30 orders of magnitude less than the $10^{47}$ molecules of SiO vapor predicted by Li2009. Thus, if there are $10^{47}$ molecules of SiO vapor in the system, they are not in equilibrium with the dust. A model of condensation of glassy silicate material in an impact produced vapor plume predicts that a significant portion of the vaporized material does not condense during its initial expansion into space (Johnson and Melosh 2012a, Johnson Melosh 2012b). Thus, the presence of abundant SiO vapor, which is out of equilibrium and still condensing onto amorphous silica dust, is an expected outcome for material condensed from an impact produced vapor plume. This process requires impact velocities higher than ~15 km s$^{-1}$ for a non-porus silica impactor and target with lower impact velocities required for less refractory and/or more porous materials (Melosh 2007, Johnson and Melosh 2012a).

## 4.2 – Timescale for Condensation

As discussed in **Section 3.2**, Dynamical modeling, of the <u>circumstellar</u> debris disk created by the Moon forming impact, shows that typical collision velocities in the disk are $\geq$ 5 km s$^{-1}$ (Jackson and Wyatt 2012). These high collisional velocities are a product of the inclined and eccentric orbits of the dust particles that are scattered when making close encounters with Earth. Assuming a similar sized planet is scattering particles in the HD 172555 debris torus, we expect that particles should collide with velocities of $v_{col} \geq$ 5 km s$^{-1}$. Although vapor created in an initial hypervelocity impact starts with an initial eccentricity and inclination, we expect that any initial eccentricity or inclination should be quickly damped down. Thus dust particles will initially move through the vapor with relative velocities of $v_{col}$. The initially large relative velocity of dust particles with respect to the vapor will decay in a time $\tau_{damp}$, such that the dust particle hits a mass of vapor equivalent to its mass, or

$$M_{dust} = \frac{4}{3}\pi a^3 \rho = m \, v_{col} \, n_{SiO} \, \pi a^2 \, \tau_{damp} \, , \qquad (11)$$

where $\rho = 3000$ kg m$^{-3}$ is the assumed density of the dust, $v_{col} = 5000$ m s$^{-1}$ is the assumed collision velocity, $a$ is the radius of the dust, and all other variables have been previously defined. Solving equation 11 using the estimate of $10^{47}$ molecules of SiO made by Li2009 we find

$$\tau_{damp} = \frac{4\,a\,\rho}{3\,m\,v_{col}\,n_{SiO}} \approx 1 \left(\frac{a}{\mu m}\right)\left(\frac{r}{AU}\right)^2 \text{yr} \,. \qquad (12)$$

For submicron dust, any inclination or eccentricity causing a relative velocity with respect to the vapor should decay in less than a year. We expect that most of the vapor will condense onto the submicron dust, which dominates the surface area of the debris (**Section 3.2**). Because this dust has little or no relative velocity with respect to the vapor, we expect that condensation will occur primarily by diffusion. Assuming a sticking coefficient of unity, we can estimate the time it takes the vapor to condense as

$$\tau_{cond} = \frac{N_{SiO}}{SA_{dust}\,\Phi} \qquad (13)$$

where $N_{SiO}$ is the number of SiO molecules and $SA_{dust}$ is the surface area of the dust particles. Then $\Phi$ is the molecular flux given by $\Phi = n_{SiO}v_{th}/4$, where $n_{SiO}$ is the number density of SiO given by $n_{SiO} = N_{SiO}/V_{torus}$, and the thermal velocity $v_{th} \approx 360$ m s$^{-1}$ at $T = 340$ K. Li2009 estimated the surface area of the fine dust to be $SA_{dust} = 4 \times 10^{21} - 2 \times 10^{22}$ m$^2$. Combining all of this we find

$$\tau_{cond} = \frac{4 \, V_{torus}}{SA_{dust} \, v_{th}} \approx (10^3 - 10^4) \left(\frac{r}{AU}\right)^2 \text{yr} \, . \qquad (14)$$

This result is independent of the total amount of SiO vapor in the system. This estimate also assumes that the SiO vapor is at the same temperature, ~340K, as the warmest fine dust. This assumption is likely a lower limit on the SiO vapor temperature, making our lifetime estimate a maximum condensation lifetime. Thus, if we are observing HD 172555 within $10^3$-$10^4$ yr of the initial impact that created the dust torus, we cannot rule out the existence of SiO vapor using arguments based on the condensation timescale.

## 4.3 – Timescale for SiO Photo-destruction

The long condensation lifetime of SiO vapor in HD 172555 allows us to treat material in the vapor phase and dust as a completely separate chemical system, simplifying the chemistry substantially. Although SiO vapor has a long condensation lifetime, we must also estimate the rate of the removal of SiO via the photo-dissociation reaction SiO + hν → Si + O. The flux of photons able to dissociate the SiO molecule at 6 AU from the HD 172555 primary is given by

$$\Phi_{(\lambda < \lambda_0 = 1500 \, \text{Å})} = \int_0^{\lambda_o} F_\lambda \frac{\lambda}{hc} \, d\lambda \approx 1.5 \times 10^{16} \, m^{-2} s^{-1} \qquad (15)$$

where $F_\lambda$ is the spectral flux density of HD 172555 at a distance of 6 AU, and $\lambda_0 = 1500$ Å is the maximum wavelength photon that can dissociate the SiO molecule (dissociation energy = 8.26 eV; Tarafdar and Dalgarno 1990). Following Ch2006, we assume that the HD 172555 primary has solar metallicity, log(g) = 4.5, and an effective temperature $T_{eff} = 8000$ K. We then estimate the stellar photospheric flux, which yields $F_\lambda$, using the appropriate stellar photospheric emission model (Kurucz 1992). Combining this, we find the value of $\Phi_{(\lambda < \lambda_0 = 1500\,\text{Å})} \approx 1.5 \times 10^{16}\ m^{-2} s^{-1}$ quoted in equation 15.

Following Tarafdar and Dalgarno (1990) we use $\sigma_{(\lambda < \lambda_0 = 1500\,\text{Å})} \approx 2 \times 10^{-21}$ m$^2$ to estimate the SiO photo-dissociation cross-section at $\lambda \leq 1500$ Å. The lifetime of a free SiO molecule at 6 AU from the primary is then given by

$$\tau_{\text{dis}} = \frac{1}{\Phi_{(\lambda < \lambda_0 = 1500\,\text{Å})}\ \sigma_{(\lambda < \lambda_0 = 1500\,\text{Å})}} = 0.01 \ \text{yr}. \quad (16)$$

Thus, if SiO is not optically thick at $\lambda \leq 1500$ Å, it will be destroyed in less than 0.01 yr by photo-dissociation. To estimate whether we expect $1.4 \times 10^{47}$ SiO molecules to be optically thick at $\lambda \leq 1500$ Å, we must compare the cross-sectional area of the SiO molecules to the cross-sectional area of the torus containing those molecules. The total cross-sectional area of $1.4 \times 10^{47}$ SiO molecules is

$$A_{SiO(\lambda < 1500\,\text{Å})} = \sigma_o N_{SiO} \approx 3 \times 10^{26} \ \text{m}^2 \qquad (17)$$

While the total cross-sectional area of the circumstellar torus is

$$A_{torus} = 2\pi R \times 2r = 1.7 \times 10^{24} \left(\frac{r}{\text{AU}}\right) \text{m}^2 \qquad (18)$$

where $R = 6$ AU. With $A_{SiO(\lambda < 1500 \text{ Å})} \gg A_{torus}$ we expect that $1.4 \times 10^{47}$ SiO molecules should be optically thick at $\lambda \leq 1500$ Å, and every photon of $\lambda \leq 1500$ Å should dissociate one SiO molecule. This increases the photo dissociation lifetime estimate to

$$\tau_{dis} = \frac{N_{SiO}}{\Phi_{dis} A_{torus}} = 0.2 \left(\frac{r}{\text{AU}}\right)^{-1} \text{yr.} \qquad (19)$$

This lifetime implies that to be observing $10^{47}$ molecules of SiO vapor, we must either be observing HD 172555 within 0.2 yrs of the original impact, which would be extremely fortuitous, or SiO is being replenished at an unrealistically high rate. This short lifetime is also inconsistent with observations of system's mid-infrared photometric flux, which has remained constant within 4% over the last 27 years (**Figure 1**).

## 4.4 – Photochemistry of a self-shielding torus

So far we have only considered the photo dissociation of SiO neglecting much of the possible chemistry of the system. In **Section 4.3** we showed that the SiO vapor will be optically thick at wavelengths $\lambda \leq 1500$ Å. In this section we consider the possibility that SiO is self-shielded against photo-dissociation. Note that "self-shielding" does not mean that there is no photolysis occuring, only that it is confined to the surface of an optically thick torus and that the re-formation back reaction creates SiO at the same rate it is destroyed.

The equilibrium chemistry of a self shielded SiO torus is dominated by four reactions. The first reaction is the photo-dissociation reaction

$$SiO + h\nu \rightarrow Si + O \qquad (20)$$

Which has a rate coefficient

$$\Gamma_{dis} = \frac{1}{\tau_{dis}} = \frac{\Phi_{(\lambda<1500\,\text{Å})}\,A_{torus}\,f_{SiO}}{N_{SiO}} \qquad (21)$$

Where $f_{SiO}$ is the fraction of the surface area of the torus occupied by SiO, as seen by photons of wavelength $\lambda < 1500$ Å. This term must be added because atomic Si vapor may also shield the SiO at $\lambda < 1500$ Å and vice versa. SiO is dissociated by photons with $\lambda < 1500$ Å while the first ionization of Si occurs at $\lambda < 1520$ Å. The two species also have very similar photo-cross-sections, with $\sigma_{SiO(\lambda<1500\,\text{Å})} = 2 \times 10^{-21}\text{m}^2$ and $\sigma_{Si(\lambda<1520\,\text{Å})} = (1-4) \times 10^{-21}\,\text{m}^2$ for wavelengths $\lambda = 912 - 1521$ Å (van Dishoeck 1988). To make order of magnitude estimates for species abundance, we approximate

$$f_{SiO} \approx \frac{N_{SiO}}{N_{Si} + N_{SiO}} \qquad (22)$$

which implies $f_{Si} = (1 - f_{SiO})$, where $f_{Si}$ is the fraction of the surface area of the torus occupied by Si, as seen by photons of wavelength $\lambda < 1500$ Å. This assumes that $f_O = 0$, where $f_O$ is the fraction of the surface area of the torus occupied by O, as seen by photons of wavelength $\lambda < 1500$ Å. This assumption is valid because O, which has a first ionization energy of ~13.62 eV requiring photons of wavelength $\lambda < 910$ Å, has a maximum value of $f_O$ given by $f_O = \Phi_{(\lambda<910\,\text{Å})}/\Phi_{(\lambda<1500\,\text{Å})} \approx 10^{-8}$. For similar reasons, we have also neglected the ionization of

SiO, which occurs at $\lambda < 1084$ Å (Tarafdar and Dalgarno 1990).  For our estimates we also approximate

$$\Phi_{(\lambda < 1500\,\text{Å})} = 1.5 \times 10^{16} \text{ photons s}^{-1}\text{m}^{-2} \approx \Phi_{(\lambda < 1520\,\text{Å})} = 2.1 \times 10^{16} \text{ photons s}^{-1}\text{m}^{-2} \qquad .$$

(23)

The second reaction we consider is the radiative association reaction

$$\text{Si} + \text{O} \rightarrow \text{SiO} + h\nu \qquad (24)$$

which occurs with a rate coefficient $k_a = 5.52 \times 10^{-24}\, T^{0.31}$ m$^3$ s$^{-1}$ (Andreazza *et al.* 1995). At a temperature of ~340 K the association rate coefficient is $k_a \sim 3.4 \times 10^{-23}$ m$^3$ s$^{-1}$.

The third reaction we consider if the photoionization of Si

$$\text{Si} + h\nu \rightarrow \text{e} + \text{Si}^+ \qquad (25)$$

which has a rate coefficient

$$\Gamma_{Ion} = \frac{\Phi_{(\lambda < 1520\,\text{Å})}\, A_{torus}(1 - f_{SiO})}{N_{Si}} \qquad (26)$$

As we will show below, Si is largely neutral and self-shielded from photoionization. Thus, we do not consider the higher ionization levels of Si.

The 4<sup>th</sup> and final final reaction we consider is the radiative recombination reaction

$$e^- + Si^+ \rightarrow Si + h\nu \qquad (27)$$

where the recombination rate $\alpha = 7 \times 10^{-18}$ m$^3$ s$^{-1}$ (Nahar 1995). The equilibrium between the photo-dissociation of SiO and its radiative association reaction gives the following

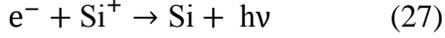

$$k_a\,[O]\,[Si] = \Gamma_{dis}\,[SiO] = \frac{\Phi_{(\lambda < 1520\,\text{Å})}\,A_{torus}\,f_{SiO}}{N_{SiO}}\,[SiO] \qquad (28)$$

where $[X] = N_X / V_{torus}$, is the concentration of species $X$ and $N_X$ is the number of species $X$ in the torus. The equilibrium between the ionization of Si and the radiative recombination gives the following

$$\Gamma_{Ion}\,[Si] = \frac{\Phi_{(\lambda < 1520\,\text{Å})} \times A_{torus} \times (1 - f_{SiO})}{N_{Si}}\,[Si] = \alpha[Si^+][e^-] \approx \alpha[Si^+]^2 \qquad (29)$$

which assumes that $[e^-] = [Si^+]$ since Si$^+$ is the only ion we have considered and we have neglected other possible sources of electrons like the stellar wind and easily ionized metal atoms. Solving these equations we find

$$N_{Si} = \frac{(1-f_{SiO})N_{SiO}}{f_{SiO}}, \qquad (30)$$

$$N_O = \frac{f_{SiO}^2 \Phi_{(\lambda < 1520\,\text{Å})} A_{torus} V_{torus}}{k_a (1-f_{SiO})} N_{SiO}, \qquad (31)$$

and

$$N_{Si^+} = \sqrt{\Phi_{(\lambda < 1520\,\text{Å})} V_{torus} A_{torus} \frac{1-f_{SiO}}{\alpha}}. \qquad (32)$$

**Figure 7** shows that if SiO vapor is produced at the same rate it is destroyed by photo-dissociation in the system, then for $10^{47}$ SiO molecules to be present, there must also be abundant atomic Si and O vapor in the system. The figure also shows that the total number of ions, atoms, and molecules in the vapor phase must be $N_{tot} = N_{Si} + N_O + N_{SiO} + N_{Si^+} > 5 \times 10^{48}$ corresponding to more than $3 \times 10^{23}$ kg or $0.05\ M_\oplus$ of vapor in the system. Although it is possible to have $0.05\ M_\oplus$ of vapor in the system, this would make the HD172555 debris disk truly unique.

So far we have also neglected the constraints that the initial composition of vaporized material give us. In **Section 4.1** we briefly mentioned the model of Johnson and Melosh (2012a), which predicts that impact produced silicate vapor does not condense efficiently during its initial expansion into space. Assuming the vaporized material is initially composed dominantly of a combination of silica ($SiO_2$), olivine ([Mg, Fe]$_2SiO_4$), and pyroxene ([Mg, Fe]$SiO_3$), the uncondensed vapor will have $N_{Si} \leq N_O \leq 4N_{Si}$, where $N_{Si}$ is the total number of Si atoms in the system and $N_O$ is the total number of O atoms in the system. If there are no major mechanisms to remove Si or O from the system, the constraints of $N_{Si}$ and $N_O$ coupled with the results illustrated in **Figure 7**, imply that

$$N_O \sim N_{Si} \sim 10^{48} \text{ molecules.} \qquad (33)$$

So far we have neglected the possible effects of the radiation force on the vapor molecules and atoms. The ratio of the radiation force to the gravitational force $\beta$ has been calculated for atomic Si, O, and Si$^+$ in the β Pictoris system (Fernández et al. 2006). Scaling the $\beta$ values as described in **Section 3.1** we find

$$\beta_{Si} = 7.5 \pm 0.7 \text{ , } \beta_{Si^+} = 11.2 \pm 11.2 \text{ , and } \beta_O = (4.1 \pm 0.2) \times 10^{-4} \text{ .} \qquad (34)$$

For $\beta > 1/2$, the particles will quickly be removed from the system by the radiation force. Thus, atomic O will not be removed from the system by the radiation force, but Si atoms will. Using the radial velocity equation (equation 2), we estimate the time it will take Si to be removed from the system to $\tau_{F_{rad}} \approx 0.1$ yrs. The large uncertainty in $\beta_{Si^+}$ makes it impossible to estimate what effect the radiation force will have on Si$^+$.

These estimates on the effect of the radiation force are only valid if the torus is optically thin at the wavelengths corresponding to the atomic transitions. Following Hilborn (1982) and Fernández et al. (2006), the wavelength-integrated cross-section of a line transition is

$$\sigma_0 = \frac{1}{8\pi c} \left(\frac{g_k}{g_i}\right) A_{ki} \lambda_{ki}^3 \qquad (35)$$

where $g_k$ and $g_i$ are the degeneracies of levels $k$ and $i$, $A_{ki}$ is the Einstein coefficient for spontaneous emission from level $k$ to $i$, and $\lambda_{ki}$ is the wavelength of the transition. Then the radiation force on a single atom is given by

$$F_{rad} = \frac{1}{c} \sum_{i<k} F_\lambda \, \lambda_{ki} \, \sigma_0, \qquad (36)$$

where $F_\lambda$ is the stellar flux per unit wavelength at the line center. Fernández et al. (2006) argue that the low temperatures and densities in the β Pictoris disk imply that radiative de-excitations are much faster than collisional or radiative excitations, implying that $i = 0$ or all absorptions are from the ground state. Using this assumption, and atomic data from the NIST Atomic Spectra Database, version 2.0 (Martin *et al.* 1999) we find the $\lambda_{ki} = 2514$ Å transition of Si will dominate the radiation force on Si. It is a good example of a strong transition with $A_{ki} = 7.39 \times 10^7 \ s^{-1}$, $g_k = 3$, and $g_i = 1$, with a typical $\sigma_0 = 5 \times 10^{-22}$ m². The Si atoms in the torus will be optically thick at line transition wavelengths as long as $N_{Si}\sigma_o \gg A_{torus} \approx 10^{24}$ m², or $N_{Si} \gg 2 \times 10^{45}$. Thus, if $N_{Si} \leq 2 \times 10^{45}$, Si will be quickly removed form the torus in $\tau_{F_{rad}} \approx 0.1$ yrs unless efficient braking by collisions with species with $\beta < 1/2$ occurs.

To estimate the conditions necessary for efficient braking to take place, we estimate the mean free path of vapor molecules in the torus as $\lambda = \left(\sqrt{2}\, n_{tot}\, \sigma\right)^{-1}$, where $\sigma \approx 10^{-19}$ m$^2$ is a typical molecular collisional crossection and is assumed to be approximately equal for all the vapor species and $n_{tot} = N_{tot}/V_{torus}$ is the total number density of vapor. The mean free path will be much less than the size of the torus as long as $N_{tot}(r/\text{AU}) \gg 10^{43}$. Thus, if $N_{tot}(r/\text{AU}) \gg 10^{43}$, an Si atom will undergo many collisions with particles with $\beta < 1/2$ before it can leave the torus. These collisions will act as a drag force on the Si, keeping Si atoms bound to the torus and reducing its effective Beta (Fernández et al. 2006). Thus, for the $N_O \sim N_{Si} \sim 10^{48}$ molecules required to keep $10^{47}$ molecules of SiO vapor stable against photo-dissociation, Si will not be blown out of the torus by radiation pressure.

## 5 – Solid SiO

If there are less than $10^{48}$ molecules of O and Si vapor in the system, SiO will be short lived and some other material must explain the emission feature at 8 $\mu$m, a wavelength region dominated by vibrational lines of SiO and SiO$_2$. Considering that the SiO stretch complex is centered at ~8 $\mu$m (Drira *et al.* 1997), the next most obvious candidate source is solid SiO. Hallenbeck *et al.* (1998, 2000) show that a shoulder feature at ~8 $\mu$m, similar to the feature seen in the Spitzer spectra of HD172555, naturally occurs in experiments involving the quick condensation of submicron dust grains (termed "smokes" by the authors) from MgO + SiO vapor. After a thermal annealing of the condensed silica/silicate grains, the ~8 $\mu$m feature attributed to solid SiO starts to disappear (Hallenbeck et al. 2000). However, the ~8 $\mu$m feature caused by solid SiO is very stable at temperatures below ~500 K. At 340 K it would take longer than the age of the universe to thermally anneal the grains (Hallenbeck et al. 2000). Once annealed the "smokes" are composed of the familiar species, tridymite ($SiO_2$), periclase ($MgO$), $Mg_2SiO_4$ olivine, and $MgSiO_3$ pyroxene. Since the solid SiO feature is only seen in vapor condensates, we are again lead to consider impacts with impact velocities greater than ~10 km s$^{-1}$, which have enough kinetic energy to partially vaporize bulk silicates (Melosh 2007). This alternative explanation is consistent with the conclusion made by Li2009 that the HD 172555 debris disk was created by a massive hypervelocity collision with an impact velocity greater than 10 km s$^{-1}$.

# 6 – Discussion

One timescale we have neglected in our discussion is the time is takes for an initially clumped spatial distribution of debris to become circular. In Jackson and Wyatt's (2012) numerical study of the circumstellar debris disk created by a Moon forming impact, they find the initial clumped spatial distribution of debris takes $10^4$-$10^5$ yr to become axisymmetric. The VLT lucky imaging and MIDI visibility modeling of HD 172555 implies that the debris disk/torus of HD 172555 is axisymmetric (Pantin & Di Folco 2011, Smith et al. 2012). Assuming that the $10^4$-$10^5$ orbit timescale for circularization can be scaled to the orbital timescale at 6 AU from the HD 172555 primary, we estimate it would take between $\tau_{circ} \sim 10^5 - 10^6$ yr for the HD 172555 debris disk to become axisymmetric following an initial giant hypervelocity impact. The circularization timescale $\tau_{circ} \sim 10^5 - 10^6$ yr is at least 10 times longer than the maximum condensation lifetime of SiO vapor, $\tau_{cond} \sim 10^3 - 10^4$ yr (**Section 4.2**). If $\tau_{circ} \sim 10^5 - 10^6$ yr is an accurate estimate for the minimum age of the HD 172555 debris disk, then all of the SiO vapor created in the initial massive impact will have condensed onto existing grains. Either SiO gas is being continuously created, the SiO gas is all condensed into an amorphous solid phase, or simple dynamical models of the system do not suffice to accurately model it. E.g., scattering of the circumstellar debris by a planet more massive than Earth could cause the HD 172555 debris disk to become axisymmetric on a shorter timescale than the estimates of Jackson and Wyatt's (2012). It is apparent that more detailed dynamical modeling of the HD 172555 debris disk is needed.

The Herschel GASPS key program detections of the atomic O 63 $\mu$m line in HD 172555 estimates an atomic oxygen mass of $2.5 \times 10^{-2} \, R \, M_{\oplus}$, where $R$ is the distance of the oxygen to the star in AU (Riviere-Marichalar et al., astro-ph 2012). Assuming the oxygen is co-spatial with the dust are $R \approx 6$ AU, this corresponds an oxygen mass of $0.9 \, M_{\oplus}$. In Section 4.4 we estimate the abundant Si and O vapor, ~$10^{48}$ atoms or $0.05 \, M_{\oplus}$, are required to keep SiO vapor from being destroyed by photo-dissociation in less than 0.2 yr. Thus, the observations of Riviere-Marichalar et al. (2012) show that there is more than enough oxygen in the system to keep $10^{47}$ molecules of SiO vapor stable against photo-dissociation. However, if follow up observations do not detect SiO vapor the 8 $\mu$m feature has to be explained by some previously overlooked mechanisms, the most likely of which being emission from solid state SiO.

The mid-infrared lucky imaging of Pantin & Di Folco (2011) and the imaging and mid-infrared interferometry of Smith *et al.* (2012) suggest that there may be some (~0.5 to 1 AU) separation of the 10 and 20 um emitting regions, with the longer wavelength emission coming from regions farther from the central star. If true, this could act as a verification of the constant hypervelocity grinding model: smaller, hotter dust, producing the bulk of the 10 $\mu$m radiation, is concentrated near the collision region, where the nodes of the rubble orbits intersect, while the large, dark, cold rubble first released by the initial massive planetesimal-planetesimal impact dominates the emission outside of the interaction region, near the apoapsis.

## 7 – Conclusion

We find that the observed infrared excess of HD 172555 corresponds to a dusty disk or torus, which is located at ~6 AU from the primary and was created in a recent planetary scale hypervelocity impact. Our new spectral modeling work and calculations of the radiation pressure on fine dust in HD 172555 provide a self-consistent explanation for the apparent contradiction of a radiation pressure blowout lifetime less than one year for submicron grains and an observed mid-infrared photometry, dominated by submicron grains, that has been stable for at least 27 years. The fine dust smaller than ~0.1 µm is dominantly composed of amorphous silica and strongly decouples from the radiation field so that it will not be lost from the system by radiation pressure blowout. We conclude that in HD 172555, dust between 0.1 µm and 1.5 µm is quickly blown out of the system by radiation pressure but the mid-infrared emission, dominated by dust smaller than 0.1 µm, and larger than 1.5 µm, can remain stable on $10^5$ yr time scales if limited by Poynting Robertson drag and even longer if fine dust is replenished by subsequent hypervelocity grinding.

We also explore the unconfirmed claim, made by Li2009, that ~$10^{47}$ molecules of SiO vapor are needed to explain a strong emission feature at ~8 µm in the Spitzer IRS spectrum of HD 172555. We find that SiO vapor cannot be in equilibrium with the dust and should condense in $10^3$-$10^4$ yrs, a timescale much shorter than the dynamical evolution time required to circularize the system. Additionally, unless there are ~$10^{48}$ atoms or 0.05 $M_{\oplus}$ of Si and O vapor in the system,

SiO vapor should be destroyed by photo-dissociation in less than 0.2 yr. A new detection of massive amounts of OI in the system by Herschel suggests that this might be the case. Motivated by the unique conditions required to keep SiO vapor in HD 172555, we have considered an alternative explanation, that the 8 μm feature is produced by solid SiO. Experiments by Hallenbeck *et al.* (1998, 2000) show an ~8 μm feature associated with solid SiO naturally occurs in fine grained silicate "smokes" created by quickly condensing vaporized silicate. Thus, both the SiO vapor and solid SiO explanations for the ~8 μm feature require vaporization of silicate material, which requires impact velocities higher than ~10 km s$^{-1}$ (Melosh 2007). Thus, when other vaporization mechanism can be ruled out, we conclude that the ~8 μm feature is diagnostic of dust generated by a hypervelocity impact with an impact velocity greater than ~10 km s$^{-1}$, consistent with Li2009.

## Acknowledgements:

This research was supported by NASA PGG grant NNX10AU88G. We also thank the anonymous referee for their useful comments.

## Appendix A- Rejected models

### A.1 - Magma Planet / Hot Disk Emission.

One of the most exciting possible sources of the observed mid-IR emission we have considered is the direct detection of emission from a magma ocean covered world. The kilometers-deep ocean could easily be created in the giant impacts required to form rocky worlds of Earth-size or larger. However, we can rule out this possibility, as it would be short lived and is unable to produce the total luminosity observed by Spitzer. In this section we also consider the possibility of a self-luminous circumplanetary magma disk. Another way of saying self-luminous is that the planet or disk is not in thermal equilibrium with the central star. These systems are attractive because the fine dust and vapor would naturally be stable as long as the planet or disk is still hot. We can estimate the lifetime of such a system by comparing the total amount of energy an impact brings into the system to the rate at which the system is radiating energy.

The total luminosity implied by the IR excess seen in HD 172555 is $2 \times 10^{24}$ W (Li2009). We can then estimate the lifetime of the system as shown below.

$$\tau \sim \frac{\text{energy from impact}}{\text{Observed radiation power}} \sim \frac{\frac{1}{2} M_{imp} V_{imp}^2}{2 \times 10^{24} \text{ W}} \qquad (A1)$$

Even if we assume an extremely high impact velocity of ~50 km s$^{-1}$ and an impactor mass, $M_{imp} = 6 \times 10^{23}$ kg, or $0.1 M_{\oplus}$, we find $\tau \sim 24$ yr. We know that IR photometry of HD 172555 has been stable since its IRAS discovery in 1983 to its re-observation by WISE in 2010 implying a minimum lifetime of ~27 yr (**Figure 1**). This minimum lifetime is a strong argument against all but the most extreme self-luminous sources.

In addition to the estimated lifetime of such a system, we can also estimate the size and temperature of such a system required to radiate energy at the rate we observe. We assume that the body radiates as a black body and find,

$$2\times10^{24}\text{W} = 4\pi R^2 \sigma T^4, \qquad \text{(A2)}$$

Where R is the radius of the body, $\sigma$ is the Stefan-Boltzmann constant, and $T$ is the temperature of the body. The black body temperature of the disk of a magma ocean planet is probably less than $T\sim3000$ K (Miller-Ricci *et al.* 2009). This means such a system would have to have a physical radius, $R > 2 \times 10^5$ km. This size is several orders of magnitude larger than any known planet therefore ruling out the possibility of magma ocean planet. This large physical size is still consistent with a circumplanetary disk, but the temperature we used $T\sim3000$ K is much higher than the temperature implied by spectroscopic observations. If we instead assume the black body temperature of the disk is ~200 K, or thermal equilibrium at 6 AU as shown by Li2009, we find the disk must have a radius, $R > 2 \times 10^7$ km. As we will show in the next section this size disk is consistent with the size of the Hill sphere of a Jovian planet located at ~6 AU from the primary.

We have ruled out a magma ocean planet as a possible source for the IR excess because such a body could not possibly radiate enough energy to explain the flux seen by Spitzer, even at temperatures much higher than the dust temperature implied by the Spitzer spectra. In addition to ruling out a magma ocean planet as a possible source, the short lifetime of self-luminous source radiating thermal energy at the levels seen in HD 172555 argues against them as a possible source for the IR radiation seen by Spitzer. As such, the only systems that are consistent with observations are those in thermal equilibrium with the central star.

## A.2 - Rejected Model: Cold Circumplanetary Disk

Another system that could possibly explain the Spitzer observations of HD 172555 is a much larger circumplanetary disk whose energy budget is driven by the central star's radiation. In this system the disk is gravitationally bound to a central planet, which means the fine dust in the system can be longer lived than the radiation pressure blowout lifetime. We can estimate the size of such a disk by assuming the disk perfectly re-radiates 100% of the impinging radiation as the IR that Spitzer sees. This means that the total luminosity implied by Spitzer data is equal to the cross-sectional area of the disk multiplied by the flux from the central star at 6 AU, or

$$\frac{L_{spitzer}}{L_\star} = 5 \times 10^{-4} = \frac{\pi r_{disk}^2}{4\pi (6AU)^2} \qquad (A3)$$

and

$$r_{disk} \geq 4 \times 10^7 \text{km} \qquad (A4)$$

We can then calculate the size of a planet needed to contain a circumplanetary disk of this size in its Hill sphere. If we ignore eccentricity and assume the planet is 6 AU from a central star of mass $M_\star = 2M_\odot$.

$$M_{planet} \geq 3 \frac{R_{Hill}^3}{R_{orbit}^3} M_\star \sim 10^{27} \text{ kg} \qquad (A4)$$

This means a planet with a mass comparable to Jupiter is required to stabilize the large cold circumplanetary disk. One possibility that could lead to such a system is a large impact onto one of the moons orbiting a Jovian planet. Recent observations using the APP coronograph of VLT/NACO to search for planetary companions by Quanz *et al.* (2011) rule out a Jovian planet larger than about four Jupiter masses at a distance of 11 AU from HD 172555. The 3-sigma upper limit of this non-detection is not enough to rule out a 1 $M_{Jup}$ planet at ~6 AU, however.

Another possible source for dust in a circumplanetary disk is a tidally heated moon. Currently the Io-Jupiter system is thought to be in a quasi-steady state where ~1000 kg s$^{-1}$ leaves Io and ~1000 kg s$^{-1}$ leaves the Jovian Hill sphere (Thomas *et al.* 2004). Using the estimated age of HD 172555 as a BPMG member at ~12 Myr, and the estimated mass of dust in the system of $10^{21}$ kg (Li2009), we find the minimum rate at which an Io-like moon would have to lose mass to produce the dust seen by Spitzer is at least $10^7$ kg s$^{-1}$ (~$10^4$ times greater than any known comet or Moon in today's solar system) for $10^7$ yrs in order to produce the debris seen. $10^{21}$ kg of material is also a huge mass for a moon to lose (Ganymede, the largest moon in the solar system, is only $1.5 \times 10^{23}$ kg). Moreover, Io's mass loss is primarily ionized $SO_2$, not Silicate material (Dessler 1980, Strobel & Wolven 2001.)

Another possible source for the observed dust is motivated by work done by Canup (2010). This work argues that giant planet satellite systems form and stabilize via a series of large body accretions and subsequent migrations. The stable systems seen today are the product of a series of moon formation, migration, tidal disruption, and planetary accretion events. It is possible that at only ~12 Myr, a giant planet in HD 172555 at ~6 AU has recently (within the last ~1 Myr) completely lost a moon due to tidal disruption and breakup. We argue against this possible source using the large spatial extent the dust must have, $r_{disk} \geq 4 \times 10^7$ km. For a Jovian planet, the Roche limit is on the order of $R_{roche} \sim 10^5$ km. The Roche limit scales as the radius of the planet, so we would need a planet more massive than the Sun to have a tidal disruption create a disk of the required radius. It is remotely possible that through the process of angular momentum transport, a small fraction of this material is transported to larger radii but analysis that is more detailed is required to determine if this idea is plausible.

Although some of the circumplanetary disk models may be able to explain the spectroscopic observations of HD 172555, tentative observations of the system including both interferometric visibility and photometric imaging are inconsistent with a localized dust source (Smith et al 2012, Pantin & Di Folco 2011). This rules out all of the circumplanetary disk models, unless we consider a leaky Hill sphere scenario. In this scenario material is only weakly bound to the Jovian Hill sphere and is constantly being lost to the circumstellar torus. This scenario does not change any of the analysis presented in the main text, it only allows for a somewhat different and arguably less probable creation scenario for the circumstellar debris disk or torus of HD 172555.

**Figures**

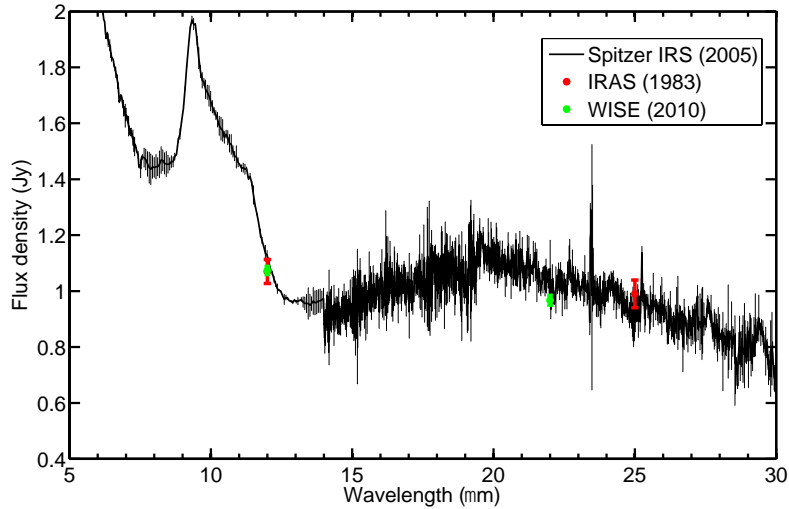

**Figure 1:** Infrared photometry and spectrum of HD 1725555. The black curve is the Spitzer IRS spectrum (observation in 2005) with 1-$\sigma$ uncertainty plotted as thinner vertical bars (Ch2006). The strong emission peak at ~9 μm, interpreted as emission from amorphous $SiO_2$, is apparent. The red points, with 1-$\sigma$ uncertainty plotted, correspond to color corrected flux density from IRAS Faint Source Catalogue v2.0 (observation in 1983) in the 12 μm and 25 μm band (Neugebauer et al. 1984, Moshir et al. 1992). The green points, with 1-$\sigma$ uncertainty plotted, correspond to color corrected-flux densities from WISE All-Sky Source Catalogue (observation in 2010) in the 12 μm and 22 μm band (Wright et al. 2010). For the 12 μm bands we used the appropriate color correction for a black body spectral energy distribution characterized by a temperature of 245 K. For the 22 μm and 25 μm bands we used the appropriate color correction for a spectral energy distribution with $F_\nu \propto \nu^{-1.5}$ (Wright et al. 2010, Beichman et al. 1998). The figure illustrates that the infrared flux from HD 172555 has remained fairly constant, within ~4% over 27 years.

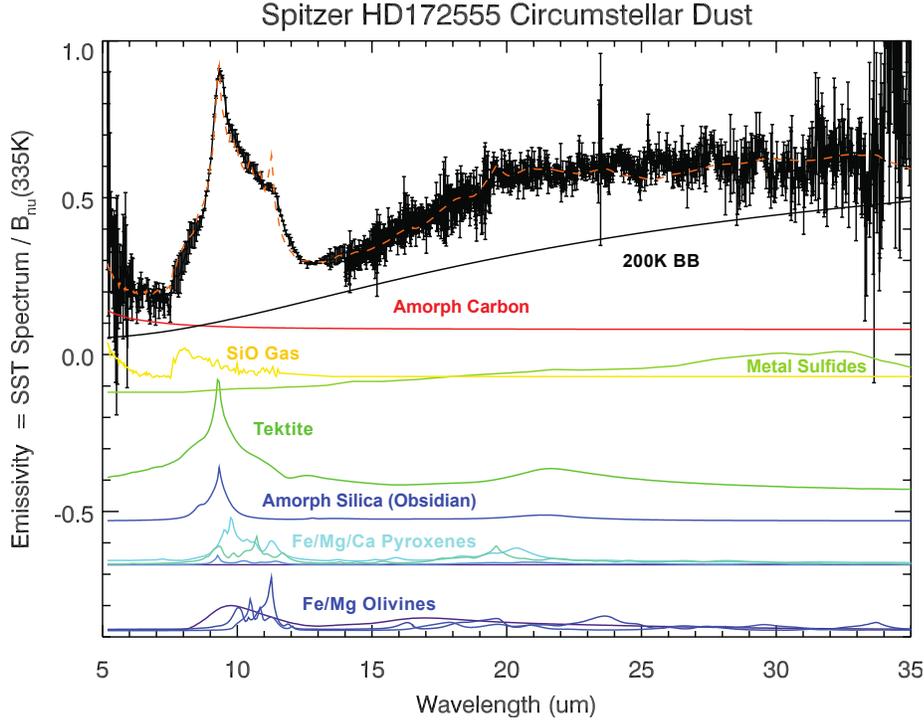

**Figure 2:** The black data with error bars is the Spitzer IRS emissivity spectrum of HD172555. The orange dashed line is emissivity spectrum of the best-fit compositional model. Relative abundances for the best-fit model can be found in Li2009. The best-fit model uses small, solid, optically thin dust grains "fine dust", SiO gas, and a population of large, cold dust grains "rubble" (200 K blackbody). The amplitude of each emissivity spectrum gives the relative contribution of each species to the total observed flux. A 335 K blackbody temperature dependence has been divided into the as-observed flux to create the emissivity spectrum. For more information on spectral modeling see **Section 3.2** and/or Li2009. (After Li2009).

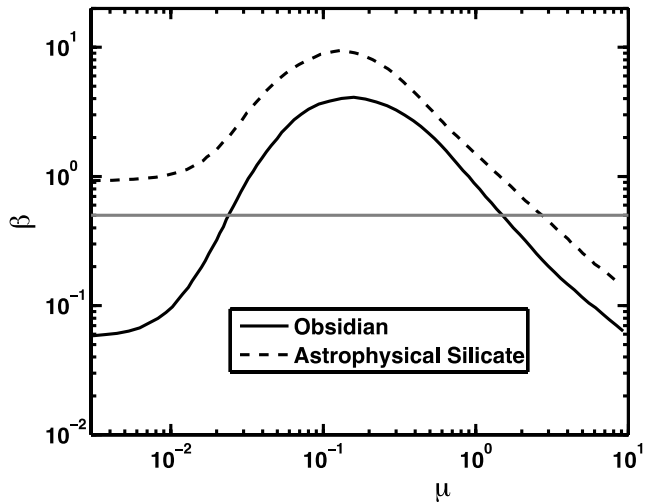

**Figure 3:** The ratio of the radiation force to gravitational force, β, as a function of particle size for obsidian (solid black line) and astrophysical silicate (dashed black line) dust orbiting HD 172555. The calculation of β assumes the particles are ~75% $SiO_2$ and ~14% $Al_2O_3$ by mole (Artymowicz 1988 and references therein) while the spectral modeling of Li2009 shows that the "fine dust" is dominated by a combination of Bediasite and Obsidian with an average composition that is ~69% $SiO_2$, 10% MgO, and 9% $Al_2O_3$. The horizontal solid line acts as a guide to the eye to show where $\beta = 1/2$, illustrating that obsidian particles between 0.02 µm and 1.5 µm should be quickly ejected from the system by radiation pressure. For astrophysical silicate, all particles smaller than 2.7 µm should be ejected from the system by radiation pressure.

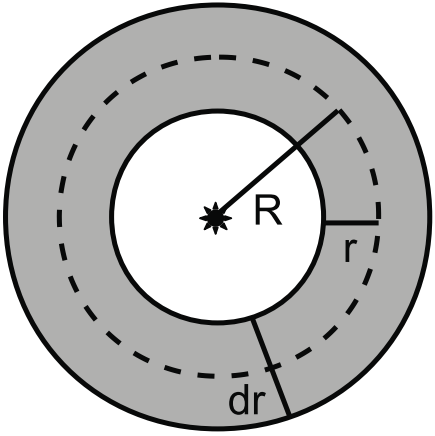

**Figure 4:** Schematic showing the geometry of the debris disk or torus. $R$ is the distance of the torus from the star, $r$ is the minor radius of the torus, and $dr$ is the disk or torus width. Note that $r = dr/2$.

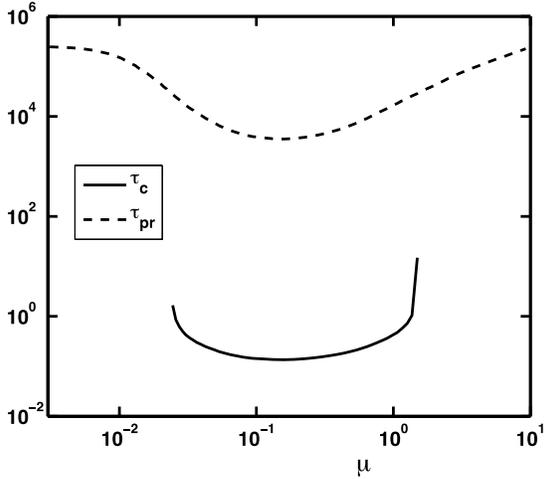

**Figure 5:** Particle lifetime estimates, plotted as a function of particle size for obsidian dust at 6 AU in HD 172555. The solid line represents estimates of the crossing time $\tau_c$ of particles blown out by radiation pressure, assuming a torus with minor radius $r = 1$ AU. The dashed line represents estimates of the lifetime of particles, spiraling into the primary from an original orbit of $R = 6$ AU due to Poynting Roberston drag given by $\tau_{pr} = (cR^2/2GM_\star\beta)$, where $M_\star = 2\ M_\odot$ (Burns *et al.* 1979).

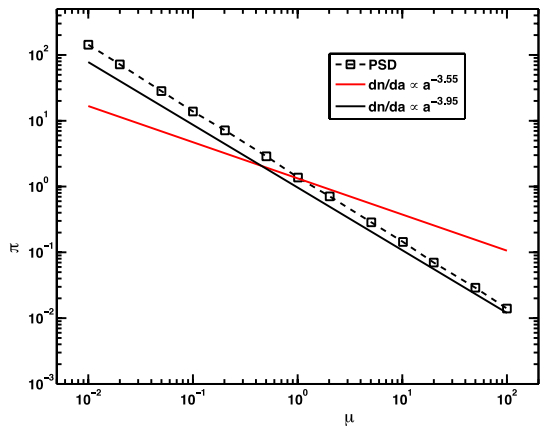

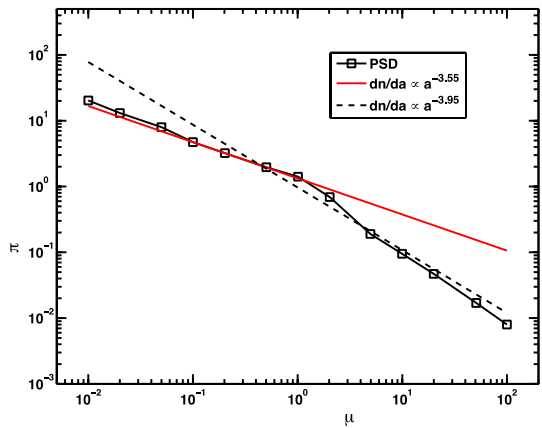

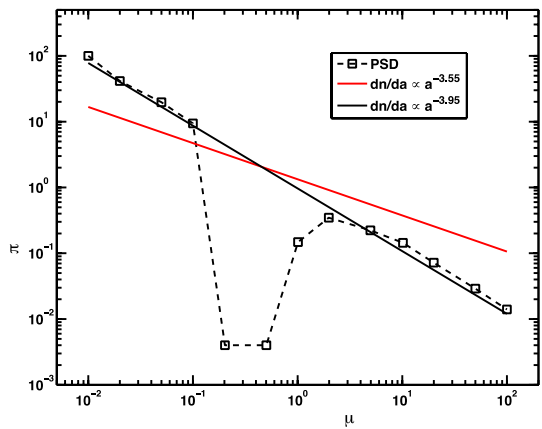

**Figure 6 - Allowable Particle Size Distributions (PSDs)** from the best-fit spectral models to the Spitzer HD 172555 5-35 μm IRS spectrum. Squares on the dashed lines represent each logarithmic bin in the PSDs used to fit the Spitzer IRS spectrum. The solid red line and solid

black line act as guides to the eye and are described in the legend. **Top:** single power law $dn/da \propto a^{-3.95\pm0.10}$ as published in Li2009. **Middle:** 2-part broken power law, with $dn/da \propto a^{-3.55}$ below 1 μm and $dn/da \propto a^{-4.02}$ above 1 μm. **Bottom:** $dn/da \propto a^{-3.95}$ with a reduction in the number of grains between 0.1 μm and 1 μm that is consistent with grains being removed by radiation pressure blowout.

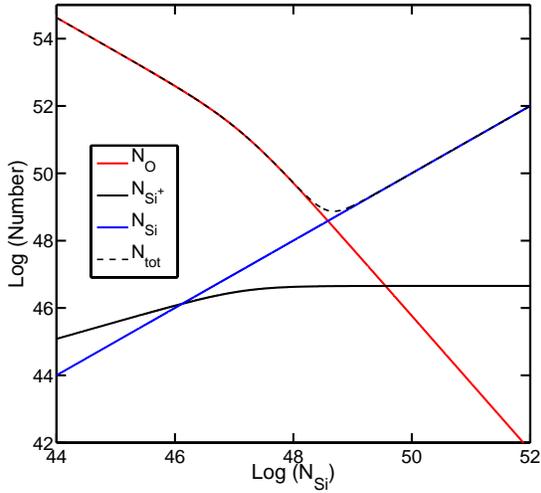

**Figure 7**: Equilibrium chemistry of HD 172555 torus. The number of various species in the torus is plotted as a function of the number of Si atoms $N_{Si}$ in the system. The curves are described in the legend and $N_{tot} = N_{Si} + N_O + N_{SiO} + N_{Si}$. The calculation assumes $N_{SiO} = 1.4 \times 10^{47}$, $A_{torus} = 2\pi R \times 2r = 1.7 \times 10^{24}$ m$^2$, $V_{torus} = 2\pi R \times \pi r^2 = 4 \times 10^{35}$ m$^3$, $\Phi_{\lambda<1520\,\text{Å}} = 2.1 \times 10^{16}\text{s}^{-1}\text{m}^{-2}$, and $r = 1$ AU.